\documentclass[aps,prl, twocolumn, superscriptaddress]{revtex4-1}

\usepackage{graphicx}
\usepackage{dcolumn}
\usepackage{bm}

\begin{document}


\title{Collapse modes in SC and BCC arrangements of elastic beads}




\author{Igor A. Ostanin}
\email{i.ostanin@utwente.nl}
\altaffiliation[Also at ]{Skolkovo Institute of Science and Technology, Skolkovo Innovation Center, 3 Nobel Street, Moscow 121205, Russia}
\affiliation{Multi-Scale Mechanics (MSM), Faculty of Engineering Technology, CSMM, MESA+, University of Twente, P.O. Box 217, 7500 AE Enschede, The Netherlands
}%

\author{Artem R. Oganov}

\affiliation{%
	Skolkovo Institute of Science and Technology, Skolkovo Innovation Center, 3 Nobel Street, Moscow 121205, Russia
}%

\author{Vanessa Magnanimo}%
\altaffiliation[Also at ]{Construction Management \& Engineering (CME), Faculty of Engineering Technology, MESA+, University of Twente, P.O. Box 217, 7500 AE Enschede, The Netherlands}
\affiliation{Multi-Scale Mechanics (MSM), Faculty of Engineering Technology, CSMM, MESA+, University of Twente, P.O. Box 217, 7500 AE Enschede, The Netherlands
}%

\date{\today}

\begin{abstract}
Collapse modes in compressed simple cubic (SC) and body-centered cubic (BCC) periodic arrangements of elastic frictionless beads were studied numerically using the discrete element method. Under pure hydrostatic compression, the SC arrangement tends to transform into a defective hexagonal close-packed or amorphous structure. The BCC assembly exhibits several modes of collapse, one of which, identified as cI16 structure, is consistent with the behavior of BCC metals Li and Na under high pressure. The presence of a deviatoric stress leads to the transformation of the BCC structure into face-centered cubic (FCC) one via the Bain path. The observed effects provide important insights on the origins of mechanical behavior of atomic systems, while the elastic spheres model used in our work can become a useful paradigm, expanding the capabilities of a hard sphere model widely used in many branches of science. 
\end{abstract}


\maketitle


It has been known since long ago that monodisperse spherical elastic particles enclosed in a confined space tend to form regular arrangements: \textit{e.g.}, depending on the boundary conditions, they can form either face-centered cubic (FCC) or hexagonal close-packed (HCP) structures. These arrangements of elastic spheres remain stable under compressive loads with significant deviatoric stress components, which is often used for storage of spherical goods. Simple cubic (SC) and body-centered cubic (BCC) structures of monodisperse spherical particles are unstable under compression, although they can be stabilized by frictional forces. 

Packing of elastic spheres under pressure is one of the simplest models of the atomic-scale structure of a pressurized material. Atomic BCC and especially SC structures have much lower densities than FCC and HCP, and could be expected to disappear under pressure. Reality, however, is much more complex and interesting. SC is the simplest possible structure, and, as nature often favors simplicity, could be expected to be common; however, at normal conditions it is found only in the alpha form of polonium, and even there it is stabilized by relativistic effects \cite{Legut,Min}. The much more common BCC structure is observed in many metals; perhaps surprisingly, it is often stable at very high pressures, and the densities of BCC phases of metals are only marginally lower than those of FCC or HCP phases, in contrast with the substantial 6.7\% difference predicted by the hard sphere model. 

In this Letter, we explore the modes of collapse for SC and BCC arrangements of elastic frictionless beads and compare our observations with the physics of real atomic structures. The collapse of these arrangements leads to new packings with a surprising variety of possible morphologies, including amorphous structures, defect-free or defective crystalline structures, and mixed-type arrangements. Under certain conditions, the BCC packing of beads exhibits a transition to the cI16 structure, similar to some BCC metals under high pressure \cite{Nature_lithium, PNAS_sodium}. The presence of deviatoric stress leads to the seamless transformation of BCC structure to FCC via the so-called Bain path. 

\paragraph{Method.}--- We used the discrete element method \cite{Cundall_Geo_1979} to study the rearrangements of elastic beads, employing the open source DEM package YADE \cite{Yade} in the calculations. The damped dynamics of equal-sized rigid spherical beads with mass $M$, radius $R$, volume $V\mathrm{_p} = \frac{4}{3} \pi R^3$ was computed using the velocity Verlet time integration scheme. The  interaction of the beads is described by a contact models that links the overlap $\delta$ and the intercenter repulsion force $F$. We used the Hertzian contact model, that provides the exact solution for the force-displacement relation in the case of identical elastic frictionless spheres:
\begin{equation}
F = \frac{4}{3} Y^{*} R^{* \frac{1}{2}} \delta^{\frac{3}{2}},
\end{equation} 
where $R^{*} = \frac{R}{2}$ is the effective contact radius, $Y^{*} = \frac{Y}{2(1-\nu^2)}$ is effective Young's modulus, $Y$ and $\nu$ are Young's modulus and Poisson's ratio of the particle material, respectively. Because the frictionless contact model does not apply torques or shear forces at a contact point, there are no rotations in our simulations.

In our experiments, a supercell containing $N \times N \times N$ periodic cells of the crystalline (SC or BCC) arrangement of $K$ identical spherical particles was confined within a cubic box with periodic boundary conditions. At the beginning of the simulation, the assembly was undeformed. Then, strain contraction $(\epsilon_x, \epsilon_y, \epsilon_z)$ was gradually applied by adjusting the $x, y,$ and $z$ sizes of the periodic cell while the homogenized compressive stress components $\sigma_x, \sigma_y, \sigma_z$ were measured (this simulation uses the standard periodic triaxial controller from YADE suite \cite{Yade}). During the simulation, we monitored the intermediate configurations of beads, homogenized principal stresses and strains, cell volume $V\mathrm{_c}$, elastic strain energy per particle $E$, and density of assembly $\phi = K V\mathrm{_p} / V\mathrm{_c}$. The stability of a lattice under hydrostatic pressure $\sigma_0$ is also characterized by the enthalpy per particle $H = E + \sigma_0 \phi^{-1} V\mathrm{_p}$. The maximum strain rate was constrained to ensure a quasi-static loading with relatively small inertia terms. During the loading and equilibration, the size of the periodic cell may change significantly while the particles perform multiple rearrangements. The excess of kinetic energy is taken out by local damping, imposing forces opposite to velocities and proportional to contact forces. \cite{YADE_man} The local damping proportionality coefficient was set to $0.9$. The simulation is stopped when the goal stress is reached and unbalanced forces in the system do not exceed their tolerances ($10^{-3}$ in our simulations). The goal stress was preset as $\sigma_x= - \sigma_0+\Delta \sigma_0, \sigma_y= -\sigma_0, \sigma_z=-\sigma_0$. Here $\Delta \sigma_0$ is the magnitude of reduction of the principal stress coaligned with the $x$ axis. Unless otherwise noted, $\Delta \sigma_0 = 0$. To ensure the validity of the Hertzian contact approximation, the magnitude $\sigma_0$ did not exceed $10^{-2}Y$, guaranteeing that the value of the elastic strains in the system remains about 1\%.

\begin{figure}
	\includegraphics[width=8.5cm]{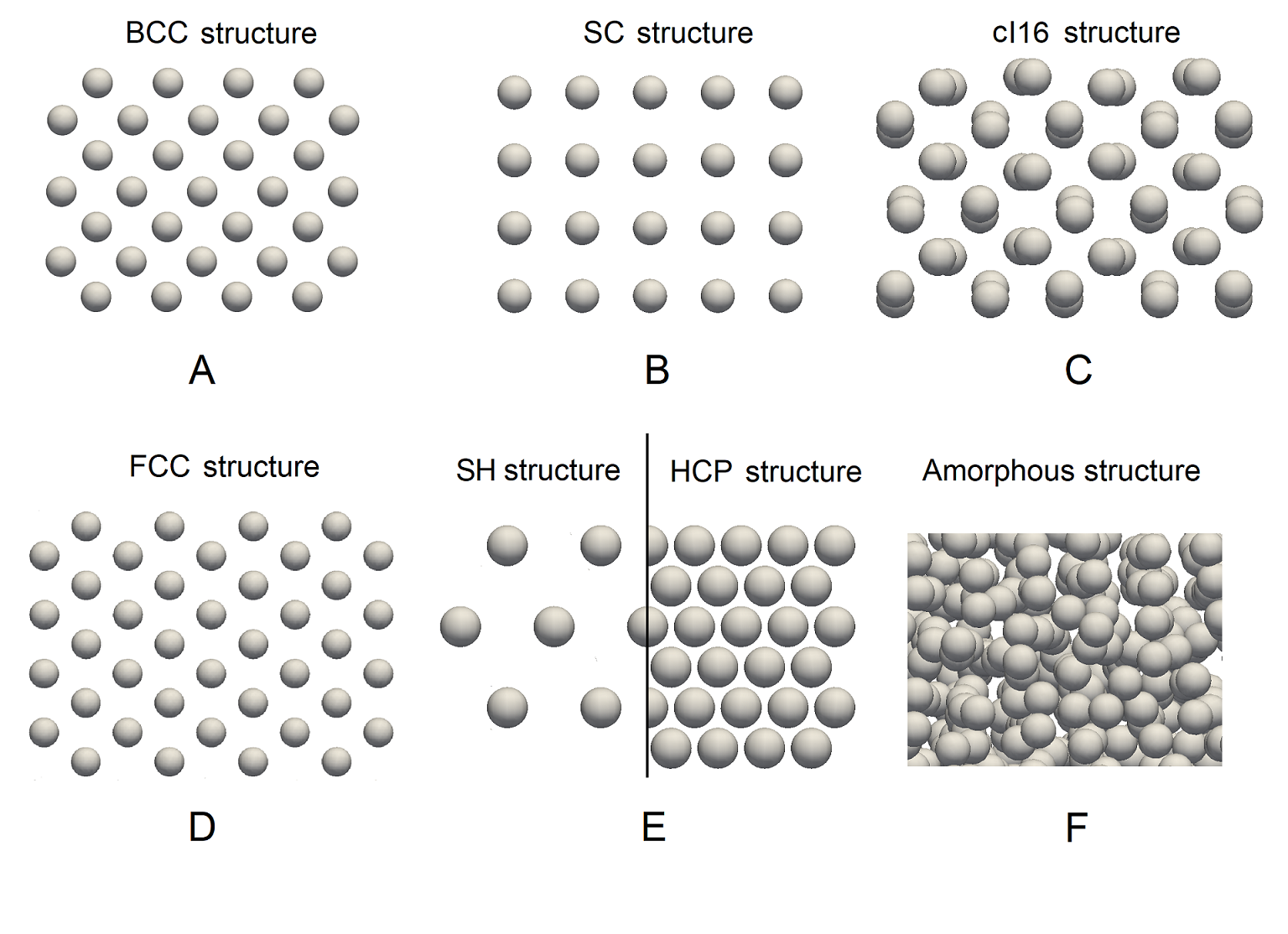}
	\protect\caption{Initial structures and their major collapse modes (side projection). For ease of perception, the beads are shown 50\% smaller than the actual beads.}
\end{figure}

\begin{figure}
	\includegraphics[width=8.5cm]{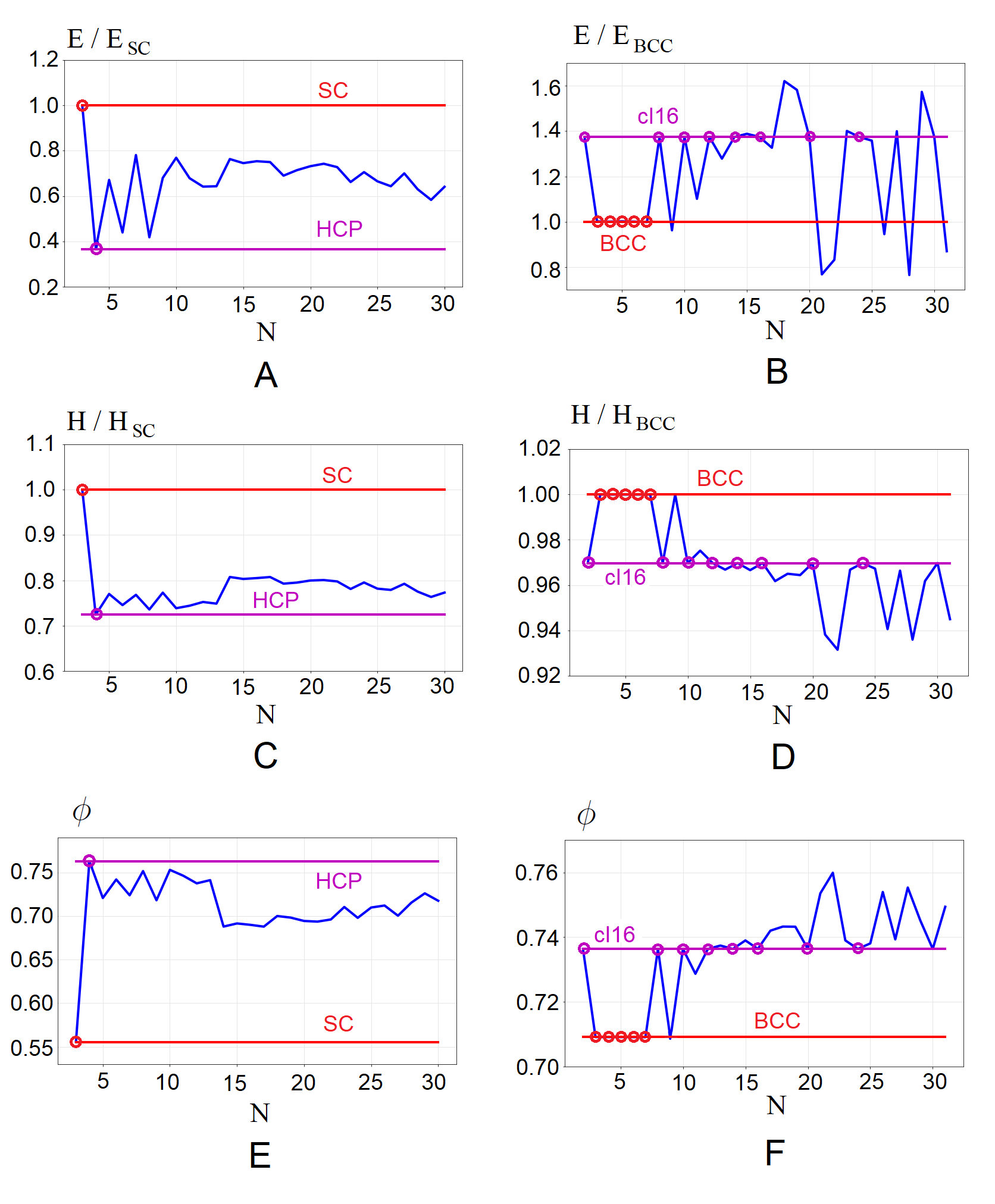}
	\protect\caption{Elastic strain energy $E$, enthalpy $H$, and density $\phi$ of (A, C, E) collapsed SC structures and (B, D, F) collapsed BCC structures as functions of the cubic supercell size $N$. Energies and enthalpies are normalized against the corresponding quantities per particle of the initial structure. }
\end{figure}

\begin{figure}
	\includegraphics[width=8.5cm]{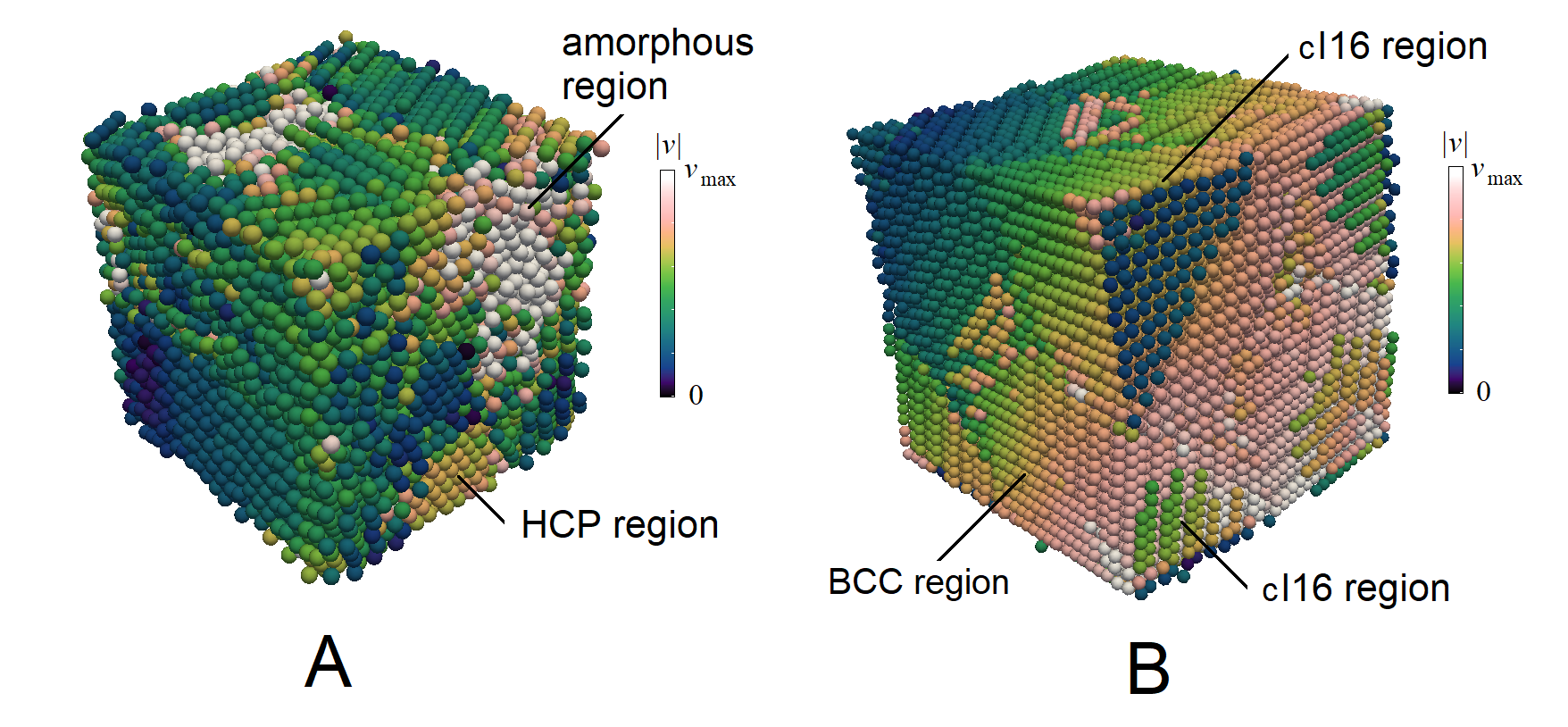}
	\protect\caption{(A) Mixed HCP/amorphous phase in large collapsed SC specimens. (B) Formation of grains in sufficiently large BCC specimens. The color legend gives the magnitude of translational velocity at the end of the simulation, helping to identify different regions.}
\end{figure}

\begin{figure}
	\includegraphics[width=8.5cm]{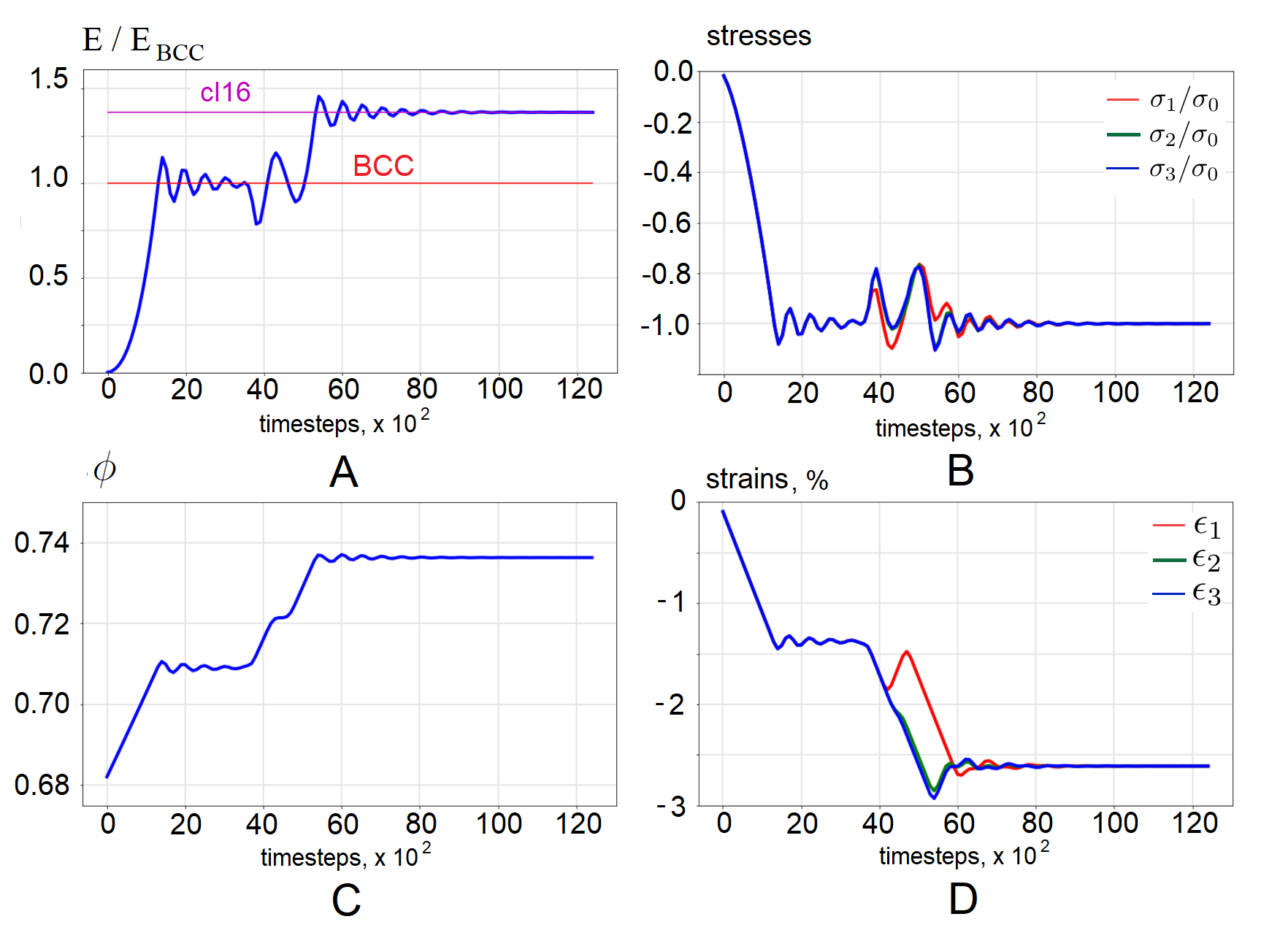}
	\protect\caption{Time evolution of (A) energy, (B) principal stresses, (C) density, and (D) principal strains during the BCC–cI16 transformation of a $12 \times 12 \times 12$ specimen.}
\end{figure}

\begin{figure}
	\includegraphics[width=8.5cm]{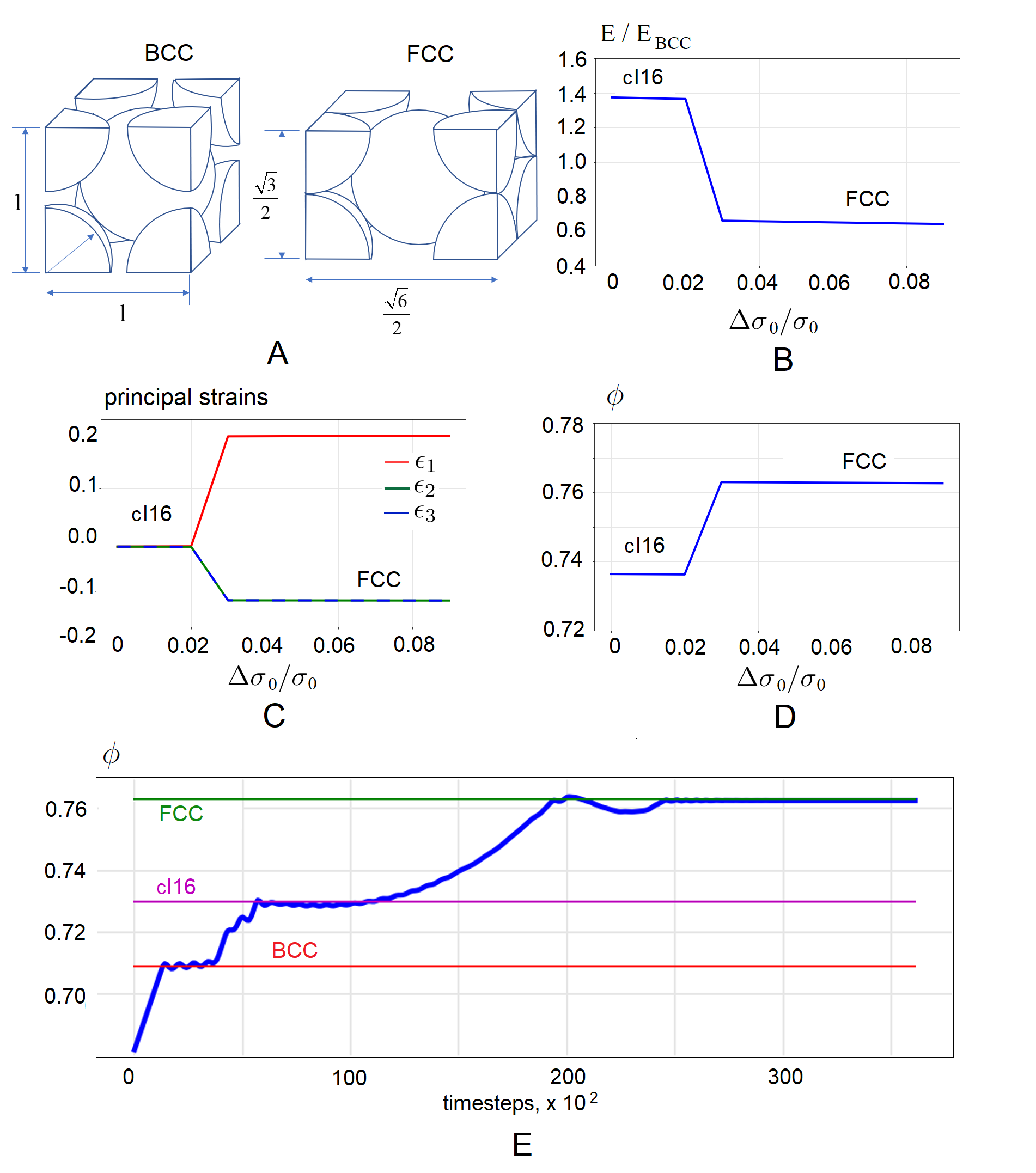}
	\protect\caption{Shift from BCC–cI16 to BCC–FCC mode of collapse in the presence of deviatoric stress in an initially BCC $12 \times 12 \times 12$ supercell. (A) Schematics of BCC–FCC transformation via the Bain path. (B) Normalized elastic strain energy per particle, (C) principal strains, and (D) density as functions of the applied reduction of first principal stress $\Delta \sigma_0 / \sigma_0$. (E) Evolution of density of packing during the transformation.}
\end{figure}

\paragraph{Results.}--- The simulations expectedly proved that the SC and BCC structures of frictionless elastic beads are unstable, unlike FCC or HCP arrangements. What is more intriguing, these two simple structures demonstrated several distinct modes of collapse (Fig.~1). 

The initial SC structure tended to collapse to either amorphous or crystalline phase, depending on the size of a specimen. The crystalline phase is usually an HCP structure. For an initially cubic supercell, the geometric constraints hinder the rearrangement of the SC structure into one with a hexagonal symmetry (although such transformation is kinetically possible, it was rarely observed in our simulations). Therefore, if the SC structure exhibits a transition into an HCP packing, structure defects are likely (see the examples in supplementary material). In this study, only the $4 \times 4 \times 4$ cubic supercell underwent a seamless transformation into a defect-free HCP structure (video 1 in the supplementary material). 

The normalized elastic strain energy, normalized enthalpy, and density of the collapsed $N \times N \times N$ SC supercell are shown in Fig.~2A,C,E. For gradually increasing $N$, two distinct phases — defective crystalline HCP and amorphous — are observed. The HCP phase is more likely for an even number of periods ($N = 4, 6, 8$). In large enough specimens, the dominant amorphous and crystalline phases coexist (Fig.~3A), whereas the energy per particle, enthalpy and relative density tend to remain nearly constant. Because the elastic contacts between spheres enable overlaps, the observed densities — $55 \%$ for the SC phase, $76 \%$ for the HCP phase, and approximately $70 \%$ for the amorphous phase — are somewhat higher than the theoretical values for rigid spheres \cite{Steinhaus_1999,Jaeger_1992} ($\pi/6 \approx 52 \%$, $ \sqrt{2} \pi / 6 \approx 74 \%$, and $64 \%$, respectively). 

The supercells of BCC structure demonstrate a richer set of collapse modes. Depending on the boundary conditions, a BCC structure can collapse into cI16 (Fig.~1C), FCC (Fig.~1D) and simple hexagonal (SH) (Fig.~1E) structures. The most likely possibility for a cubic supercell with even number of periods along every axis is the cI16 phase, which is also the only structure whose specimens are defect-free in a wide range of supercell sizes. The density of BCC structure $70.9 \%$ increases up to $73.6 \%$ for cI16 structure. 

Fig.~2B,D,F shows the elastic strain energy, enthalpy, and density of the collapsed $N \times N \times N$ BCC supercell. The BCC phase tended to form only crystalline structures. Supercells with $N= 3,4,5,6,7$ remained in the initial BCC phase (Fig.~2B), whereas those with $N=2,8,10,12,14,16,20$, and $24$ collapsed into defect-free cI16 structures. Although cI16 lattices correspond to larger stored elastic strain energies compared to BCC structures, the BCC–cI16 transition correlates with a decrease in enthalpy, which means higher stability of cI16 structures.

The BCC–cI16 transformation is a nearly coherent, although not instantaneous, rearrangement (video 2 in the supplementary material) — it takes considerable time and is accompanied by symmetry-breaking rearrangements with a dynamic evolution of system energy (Fig.~4A), principal stresses (Fig.~4B), density (Fig.~4C) and principal strains (Fig.~4D). The formation of a single cI16 cell requires $2 \times 2 \times 2$ BCC cells, therefore, in the case of an odd number of periods along each direction of a supercell, the seamless rearrangement into cI16 is impossible. Supercells with $N=11,13,15,17-19,23,25,27,29,30$ form defective cI16 structures. Supercells with $N=22, 26$ formed defective FCC structures, supercells with $N = 21, 28$ formed defective prismatic SH grains (see the supplementary material for details).  

Large specimens tend to form heterogeneous grains with distinctive grain boundaries: for example, the collapsed cI16 structure can coexist with the BCC arrangement in a neighboring grain (Fig.~3B). 

The collapse mode of a BCC arrangement, unlike that of an SC structure, appears to be sensitive to the presence of the deviatoric loading. A relatively small reduction of one of the principal stresses (approximately $3 \%$ of the baseline isotropic stress for a $12 \times 12 \times 12$ supercell) leads to another type of collapse. Instead of cI16, the BCC arrangement transforms into FCC, with a specimen expanding in the direction of the reduced stress (video 3 in the supplementary material). The transformation between the BCC and FCC structures via a tetragonal distortion is very well known as the Bain path \cite{Bain_path} (Fig.~5A). The transition corresponds to a $35 \%$ drop in stored elastic energy per particle (Fig.~5B), in contrast with the BCC–cI16 transition under hydrostatic compression, which increases the stored energy by $37 \%$.
Simple geometric considerations (Fig.~5A) show that the BCC–FCC transition in a system of rigid beads corresponds to an extension by $\sqrt{6}/2-1 \approx 0.2247$ in the direction of the reduced principal stress and a contraction by $\sqrt{3}/2-1 \approx -0.1339$ in the other principal directions. The FCC structure has 12 nearest neighbors and a density of $76.4 \%$ in the case of elastic contacts between the beads. The principal strain components (Fig.~5C) and measured density (Fig.~5D) observed in our simulations closely correspond to these predictions. A closer look at the BCC–FCC transition reveals that the supercell often goes through cI16 phase before reaching its final FCC arrangement (Fig.~5E).  

\paragraph{Discussion.}--- Our study revealed several interesting mechanical phenomena in monodisperse assemblies of elastic spheres that can be qualitatively related with the behavior of atomic structures in crystals. For example, the collapse and amorphization of an SC structure resembles the pressure-induced amorphization that occurs most often in low-density solids. \cite{Mishima_1984,Richet} Therefore, it is not surprising that here we observed it in the lowest-density SC structure. 

The most notable result of this study is the transition of the BCC structure of frictionless elastic beads under an isotropic pressure into the cI16 structure, which is similar to the rearrangement of atoms occurring in the BCC phases of lithium and sodium under extreme pressures. Hanfland et al. \cite{Nature_lithium} attributed the formation of the cI16 structure in Li to electronic structure effects. However, the cI16 phase emerges in our mechanistic model that relies on pair interactions between elastic beads, which suggests that electronic structure effects are not necessary for explaining this transition.

Recently, a similar BCC–cI16 transition has been observed in another classical model — the Monte-Carlo simulations of the BCC arrangement of classical hard spheres. \cite{Warshavsky_2018} Our study confirms and strengthens this result, demonstrating the same transition in explicit classical dynamic simulations.

Another important result of this study is the bifurcation between the BCC–cI16 and BCC–FCC modes of collapse at a certain critical deviatoric stress. The existence of such a bifurcation suggests that similar mechanisms might be found in BCC metals under high pressure with a strong deviatoric stress component. 

\paragraph{Conclusions.}--- In this work, we performed a numerical study of the dynamic evolution of unstable SC and BCC structures of frictionless elastic beads. We found that such assemblies demonstrate a surprising variety of admissible behaviors. Whereas the SC structures tend to collapse into a mixed HCP/amorphous phase, the BCC structures under pressure transform into the cI16 arrangement, similar to BCC metals lithium and sodium. It was established that the collapse of a BCC structure can follow different paths, preferring either cI16 or FCC phase, controlled by a vanishingly small change in the deviatoric stress. 
The obtained simulation results could be verified experimentally within the setup similar to \cite{Tournat2011}. However, the necessity of vanishingly small friction might constitute significant challenge for such a verification.
Our results suggest that the dissipative dynamics of classical frictionless Hertzian beads may be a meaningful and predictive model for real processes in the atomic systems. Moreover, such a model can become a useful higher-level paradigm, providing more realism than the hard sphere model, commonly used in many areas of material science.

\section{Acknowledgments}

A.R.O. thanks the Russian Science Foundation (grant 19-72-30043). I.A.O.'s postdoc project was supported by the Materials programme of the University of Twente, hosted by MESA+. The financial support from the Dutch Research Council (NWO) through the OpenMind Project “Soft Seismic Shields” is gratefully acknowledged.

\bibliographystyle{apsrev}
\bibliography{manuscript}

\begin{thebibliography}{14}
\expandafter\ifx\csname natexlab\endcsname\relax\def\natexlab#1{#1}\fi
\expandafter\ifx\csname bibnamefont\endcsname\relax
  \def\bibnamefont#1{#1}\fi
\expandafter\ifx\csname bibfnamefont\endcsname\relax
  \def\bibfnamefont#1{#1}\fi
\expandafter\ifx\csname citenamefont\endcsname\relax
  \def\citenamefont#1{#1}\fi
\expandafter\ifx\csname url\endcsname\relax
  \def\url#1{\texttt{#1}}\fi
\expandafter\ifx\csname urlprefix\endcsname\relax\def\urlprefix{URL }\fi
\providecommand{\bibinfo}[2]{#2}
\providecommand{\eprint}[2][]{\url{#2}}

\bibitem[{\citenamefont{Legut et~al.}(2007)\citenamefont{Legut, Fri\'ak, and
  \ifmmode~\check{S}\else \v{S}\fi{}ob}}]{Legut}
\bibinfo{author}{\bibfnamefont{D.}~\bibnamefont{Legut}},
  \bibinfo{author}{\bibfnamefont{M.}~\bibnamefont{Fri\'ak}}, \bibnamefont{and}
  \bibinfo{author}{\bibfnamefont{M.}~\bibnamefont{\ifmmode~\check{S}\else
  \v{S}\fi{}ob}}, \bibinfo{journal}{Phys. Rev. Lett.}
  \textbf{\bibinfo{volume}{99}}, \bibinfo{pages}{016402}
  (\bibinfo{year}{2007}),
  \urlprefix\url{https://link.aps.org/doi/10.1103/PhysRevLett.99.016402}.

\bibitem[{\citenamefont{Min et~al.}(2006)\citenamefont{Min, Shim, Park, Kim,
  Kwon, and Youn}}]{Min}
\bibinfo{author}{\bibfnamefont{B.~I.} \bibnamefont{Min}},
  \bibinfo{author}{\bibfnamefont{J.~H.} \bibnamefont{Shim}},
  \bibinfo{author}{\bibfnamefont{M.~S.} \bibnamefont{Park}},
  \bibinfo{author}{\bibfnamefont{K.}~\bibnamefont{Kim}},
  \bibinfo{author}{\bibfnamefont{S.~K.} \bibnamefont{Kwon}}, \bibnamefont{and}
  \bibinfo{author}{\bibfnamefont{S.~J.} \bibnamefont{Youn}},
  \bibinfo{journal}{Phys. Rev. B} \textbf{\bibinfo{volume}{73}},
  \bibinfo{pages}{132102} (\bibinfo{year}{2006}),
  \urlprefix\url{https://link.aps.org/doi/10.1103/PhysRevB.73.132102}.

\bibitem[{\citenamefont{Hanfland et~al.}(2000)\citenamefont{Hanfland, Syassen,
  and Christensen}}]{Nature_lithium}
\bibinfo{author}{\bibfnamefont{M.}~\bibnamefont{Hanfland}},
  \bibinfo{author}{\bibfnamefont{K.}~\bibnamefont{Syassen}}, \bibnamefont{and}
  \bibinfo{author}{\bibfnamefont{N.~e.~a.} \bibnamefont{Christensen}},
  \bibinfo{journal}{Nature} \textbf{\bibinfo{volume}{408}},
  \bibinfo{pages}{174–178} (\bibinfo{year}{2000}),
  \bibinfo{note}{doi:10.1038/35041515}.

\bibitem[{\citenamefont{McMahon et~al.}(2007)\citenamefont{McMahon, Gregoryanz,
  Lundegaard, Loa, Guillaume, Nelmes, Kleppe, Amboage, Wilhelm, and
  Jephcoat}}]{PNAS_sodium}
\bibinfo{author}{\bibfnamefont{M.~I.} \bibnamefont{McMahon}},
  \bibinfo{author}{\bibfnamefont{E.}~\bibnamefont{Gregoryanz}},
  \bibinfo{author}{\bibfnamefont{L.~F.} \bibnamefont{Lundegaard}},
  \bibinfo{author}{\bibfnamefont{I.}~\bibnamefont{Loa}},
  \bibinfo{author}{\bibfnamefont{C.}~\bibnamefont{Guillaume}},
  \bibinfo{author}{\bibfnamefont{R.~J.} \bibnamefont{Nelmes}},
  \bibinfo{author}{\bibfnamefont{A.~K.} \bibnamefont{Kleppe}},
  \bibinfo{author}{\bibfnamefont{M.}~\bibnamefont{Amboage}},
  \bibinfo{author}{\bibfnamefont{H.}~\bibnamefont{Wilhelm}}, \bibnamefont{and}
  \bibinfo{author}{\bibfnamefont{A.~P.} \bibnamefont{Jephcoat}},
  \bibinfo{journal}{Proceedings of the National Academy of Sciences}
  \textbf{\bibinfo{volume}{104}}, \bibinfo{pages}{17297}
  (\bibinfo{year}{2007}), ISSN \bibinfo{issn}{0027-8424},
  \eprint{https://www.pnas.org/content/104/44/17297.full.pdf},
  \urlprefix\url{https://www.pnas.org/content/104/44/17297}.

\bibitem[{\citenamefont{Cundall and Strack}(1979)}]{Cundall_Geo_1979}
\bibinfo{author}{\bibfnamefont{P.~A.} \bibnamefont{Cundall}} \bibnamefont{and}
  \bibinfo{author}{\bibfnamefont{O.}~\bibnamefont{Strack}},
  \bibinfo{journal}{Geotechnique} \textbf{\bibinfo{volume}{29}},
  \bibinfo{pages}{47} (\bibinfo{year}{1979}).

\bibitem[{\citenamefont{Šmilauer~et al.}(2015)}]{Yade}
\bibinfo{author}{\bibfnamefont{V.}~\bibnamefont{Šmilauer~et al.}},
  \emph{\bibinfo{title}{Yade Documentation The Yade Project.}},
  \bibinfo{edition}{2nd} ed. (\bibinfo{year}{2015}), \bibinfo{note}{dOI
  10.5281/zenodo.34073}.

\bibitem[{\citenamefont{https://yade
  dem.org/doc/formulation.html}(2020)}]{YADE_man}
\bibinfo{author}{\bibnamefont{https://yade dem.org/doc/formulation.html}}
  (\bibinfo{year}{2020}).

\bibitem[{\citenamefont{Steinhaus}(1999)}]{Steinhaus_1999}
\bibinfo{author}{\bibfnamefont{H.}~\bibnamefont{Steinhaus}},
  \emph{\bibinfo{title}{Mathematical Snapshots,}} (\bibinfo{publisher}{New
  York: Dover}, \bibinfo{year}{1999}), \bibinfo{edition}{3rd} ed.,
  \bibinfo{note}{pp. 202-203}.

\bibitem[{\citenamefont{Jaeger and Nagel}(1992)}]{Jaeger_1992}
\bibinfo{author}{\bibfnamefont{H.~M.} \bibnamefont{Jaeger}} \bibnamefont{and}
  \bibinfo{author}{\bibfnamefont{S.~R.} \bibnamefont{Nagel}},
  \bibinfo{journal}{Science} \textbf{\bibinfo{volume}{255}},
  \bibinfo{pages}{1524} (\bibinfo{year}{1992}).

\bibitem[{\citenamefont{Bain}(1924)}]{Bain_path}
\bibinfo{author}{\bibfnamefont{E.}~\bibnamefont{Bain}},
  \bibinfo{journal}{Trans. AIME} \textbf{\bibinfo{volume}{70}},
  \bibinfo{pages}{1} (\bibinfo{year}{1924}).

\bibitem[{\citenamefont{Mishima et~al.}(1984)\citenamefont{Mishima, Calvert,
  and Whalley}}]{Mishima_1984}
\bibinfo{author}{\bibfnamefont{O.}~\bibnamefont{Mishima}},
  \bibinfo{author}{\bibfnamefont{L.}~\bibnamefont{Calvert}}, \bibnamefont{and}
  \bibinfo{author}{\bibfnamefont{E.}~\bibnamefont{Whalley}},
  \bibinfo{journal}{Nature} \textbf{\bibinfo{volume}{310}},
  \bibinfo{pages}{393–395} (\bibinfo{year}{1984}), \bibinfo{note}{doi:
  10.1038/310393a0}.

\bibitem[{\citenamefont{P.}(1997)}]{Richet}
\bibinfo{author}{\bibfnamefont{R.~P. .~G.} \bibnamefont{P.}},
  \bibinfo{journal}{Eur. J. Miner.} \textbf{\bibinfo{volume}{9}},
  \bibinfo{pages}{907} (\bibinfo{year}{1997}).

\bibitem[{\citenamefont{Warshavsky et~al.}(2018)\citenamefont{Warshavsky, Ford,
  and P.A.}}]{Warshavsky_2018}
\bibinfo{author}{\bibfnamefont{V.}~\bibnamefont{Warshavsky}},
  \bibinfo{author}{\bibfnamefont{D.}~\bibnamefont{Ford}}, \bibnamefont{and}
  \bibinfo{author}{\bibfnamefont{M.}~\bibnamefont{P.A.}}, \bibinfo{journal}{J
  Chem Phys.} \textbf{\bibinfo{volume}{148}}, \bibinfo{pages}{024502}
  (\bibinfo{year}{2018}), \bibinfo{note}{doi: 10.1063/1.5009099}.

\bibitem[{\citenamefont{Merkel et~al.}(2011)\citenamefont{Merkel, Tournat, and
  Gusev}}]{Tournat2011}
\bibinfo{author}{\bibfnamefont{A.}~\bibnamefont{Merkel}},
  \bibinfo{author}{\bibfnamefont{V.}~\bibnamefont{Tournat}}, \bibnamefont{and}
  \bibinfo{author}{\bibfnamefont{V.}~\bibnamefont{Gusev}},
  \bibinfo{journal}{Phys. Rev. Lett.} \textbf{\bibinfo{volume}{107}},
  \bibinfo{pages}{225502} (\bibinfo{year}{2011}),
  \urlprefix\url{https://link.aps.org/doi/10.1103/PhysRevLett.107.225502}.

\end{thebibliography}

\end{document}